\documentclass[12pt]{article}
\usepackage{epsfig,graphicx}

\title{Relativistic theory  of string breaking in QCD}
\author{ Yu.A.Simonov\\
State Research
Center\\Institute of Theoretical and Experimental Physics, \\
Moscow, 117218 Russia}

\newcommand{\beq}{\begin{eqnarray}}
 \newcommand{\eeq}{\end{eqnarray}}
\newcommand{\be}{\begin{equation}}
 \newcommand{\ee}{\end{equation}}

\def\fun#1#2{\lower3.6pt\vbox{\baselineskip0pt\lineskip.9pt
\ialign{$\mathsurround=0pt#1\hfil ##\hfil$\crcr#2\crcr\sim\crcr}}}

\newcommand{{\SD}}{\rm SD}

\newcommand{\veb}{\mbox{\boldmath${\rm b}$}}

\newcommand{\vex}{\mbox{\boldmath${\rm x}$}}
\newcommand{\vey}{\mbox{\boldmath${\rm y}$}}
\newcommand{\ver}{\mbox{\boldmath${\rm r}$}}
\newcommand{\vesig}{\mbox{\boldmath${\rm \sigma}$}}

\newcommand{\vep}{\mbox{\boldmath${\rm p}$}}
\newcommand{\veq}{\mbox{\boldmath${\rm q}$}}

\newcommand{\vez}{\mbox{\boldmath${\rm z}$}}

\newcommand{\veL}{\mbox{\boldmath${\rm L}$}}

\newcommand{\ves}{\mbox{\boldmath${\rm s}$}}

\newcommand{\ven}{\mbox{\boldmath${\rm n}$}}
\newcommand{\veu}{\mbox{\boldmath${\rm u}$}}

\newcommand{\vew}{\mbox{\boldmath${\rm w}$}}

\newcommand{\veH}{\mbox{\boldmath${\rm H}$}}
\newcommand{\veE}{\mbox{\boldmath${\rm E}$}}

\newcommand{\veal}{\mbox{\boldmath${\rm \alpha}$}}

\newcommand{\llan}{\langle\langle}
\newcommand{\rran}{\rangle\rangle}
\newcommand{\lan}{\langle}
\newcommand{\ran}{\rangle}

\begin{document}
\maketitle
\begin{abstract}
The QCD string breaking due to quark pair creation in the vacuum confining
field, possibly  accompanied by vector, scalar  or Nambu-Goldstone bosons, is
studied nonperturbatively.  The scalar light pair creation  vertex  occurs due
to chiral symmetry breaking and has a confining form, which is computed
explicitly together with subleading vector contributions.  Dependence on light
quark mass and flavor is specifically studied. The
 dominant  scalar term   is in good agreement with the    $~^3P_0$  model and  experimental data.

\end{abstract}

\section{Introduction}
The topic of string breaking and effective decay Lagrangian has a long history.
Among the nonperturbative (np) models the most famous is the so-called $~^3P_0$
model (see \cite{1}, \cite{2} for details,  reviews and references), which can
also be written in the form of the interaction Hamiltonian for the Dirac quark
fields \be H_I = g\int d^3 x \bar \psi \psi \label{1a}\ee where $g= 2 m_q
\gamma$, $ m_q$ is the constituent quark mass and $ \gamma \approx  0.5$ is the
phenomenological parameter. In  what follows we derive effective  Lagrangian in
QCD, which
 in the relativistic invariant form
 can be written  similarly to (\ref{1a}) as \be \mathcal{L}=\int \bar \psi\mathcal{M}
\psi d^4x\label{1}\ee with $\mathcal{M}$ - calculated nonperturbatively through
string tension, while $\psi, \bar \psi$ are relativistic bispinors with current
(polar) mass. The model (\ref{1a}) is rather successful (see \cite{1,2, 3,4}
for reviews and discussion), however not deduced from the QCD Lagrangian.  In
the detailed analysis of heavy quarkonia in \cite{5} another model was
suggested, where the role of $ \mathcal{M}$ in (\ref{1}) is played by the
quark-antiquark potential $V_{Q\bar Q}(r)$ with confining  and OGE parts of
color  octet vector type. An important step was done further in \cite{4}, where
scalar and vector color singlet  potentials were carefully studied, and
similarity of  results for  $~^3P_0$ and scalar potential model was stressed.
A successful application of the $~^3P_0$ model to charmonia was done in
\cite{6} with $\gamma =0.4$, assuming harmonic oscillator wave functions. This
seemingly universal character of the model calls for the careful study of its
connection with QCD.

In the present paper we aim at the most general derivation of the string
breaking amplitude with possible accompanying radiation of $\gamma$ or
$\pi,\eta, K$, or vector and scalar mesons.

Formally the problem of string breaking or of  $q\bar q$ pair creation on the
string connecting quark $Q$ and antiquark $\bar Q$ can be reduced to the
calculation of partition function  \be Z=\int DA \exp  (\mathcal{L}_A)~
W_{Q\bar Q} (A) \det (m_q + \hat D (A))\label{A3}\ee where $ \mathcal{L}_A$ is
the standard gluonic action and $W_{Q\bar Q}(A)$ is the (external) Wilson loop
of (heavy) quarks $Q\bar Q$ with fixed contour. Using world-line representation
for  the det term  in (\ref{A3}) and retaining  only one loop of quark $q$, one
has \be Z_{\rm 1loop}= \int  DA \exp \mathcal{L}_A \left( -\frac12 tr
\int^\infty_0 \frac{ds}{s} (D^4 z) e^{-K} W_{q\bar q} (A)W_{Q\bar
Q}(A)\right).\label{A4}\ee

Thus  the  effective Lagrangian can be obtained from the averaged product of
two Wilson loops, calculated in \cite{6*}, \be \chi\equiv  \lan W_{q\bar
q}W_{Q\bar Q}\ran_A \cong \frac{1}{N^2_c} \exp (-\sigma S_\Delta).\label{A5}\ee
where $S_\Delta$ is the minimal area between contours $q\bar q$ and $Q\bar Q$.
For large masses $m_q$ and $m_Q$ one can easily reproduce the scalar potential
(sKs) model of \cite{4} with $\mathcal{L}_{\rm nonrel} = \int \sigma |\vex_q
-\vex_{\bar Q} | dt$. However in the small $m_q$ limit the  transition  from
world-line representation of the light  quark to the bispinor Lagrangian $\int
\bar \psi \mathcal{M}\psi d^4 x$ is not easy, since appearing of the scalar
operator $\mathcal{M} $ for light quark implies chiral symmetry breaking.
Therefore below we shall explicitly construct the kernel $\mathcal{M}$, which
exemplifies simultaneously confinement and spontaneous Chiral Symmetry Breaking
(CSB) for quark $q$ of vanishing mass $m_q$  in the field of  external
(possibly static for simplicity) quark $\bar Q$. When mass $m_q$ is growing, $
m_q\gg\sqrt{\sigma}$, spontaneous CSB is replaced by the explicit one, and  the
problem is much simpler.

To derive Effective String-Decay  Lagrangian (ESDL), one can use the well-known
Background Perturbation Theory (BPTh) \cite{7}, which was developed further in
\cite{8}, where the confining properties of the background were taken into
account. This theory has the prominent advantage, that it is infrared safe, and
the Landau ghost pole and IR renormalons are absent in the total perturbation
theory, where the first term is purely nonperturbative and can be calculated
via $np$  field correlators \cite{9}. The latter in their turn are calculated
selfconsistently through the only mass parameter of the theory, e.g. the string
tension $\sigma$ (see \cite{10} for a review and references).

Therefore in computing ESDL one obtains several terms. First of all the
well-known   $~^3S_1$ term \cite{11} due to gluon exchange, which in BPTh
becomes the hybrid-mediated transition, discussed in \cite{11*}, when confining
background is important (for not large energy release). Below we shall be
interested mostly in the  scalar-type terms of ESDL. As was said above, those
terms occur due to CSB in the confining background, and our derivation follows
closely previous papers \cite{12,13,14,14*}, where CSB due to confinement in
QCD was derived in great detail.

We shall specifically stress below the main mechanism, which creates for a
(massless) relativistic quark simultaneously and spontaneously confinement and
CSB, the mechanism  which  takes into account symmetry of  quark spectrum of
Hamiltonian with  scalar interaction. For massive quark of large mass the same
effect occurs due to explicit CSB. Integrating out quark degrees of freedom in
the effective Lagrangian, one obtains chiral effective Lagrangian for NG mesons
at that point of  the string, where it breaks down. In this way one obtains
ESDL with NG meson  degrees of freedom.

The mechanism, described below and in \cite{14}-\cite{14*} can be cast into
explicitly gauge invariant form (see appendix  1 of \cite{14}), which is
possible, since the systems under  consideration ($(\bar Q q)$ or ($Q\bar q)$)
are white. This is in contrast to models, considering light quark pair $(q\bar
q)$ alone.

Concluding this introduction, one should stress, that our approach to string
breaking as a CSB light-pair creation process differs in principle from the
Schwinger-type mechanism of pair creation, specific for vast electric fluxes in
QED, see \cite{17*} for discussion and references.

\section{Chiral symmetry breaking and scalar string breaking}

Consider now  the partition function of a light quark in the field of a static
antiquark  in the presence of external   currents $ v_\mu, a_\mu, s, p$
$$Z = \int DA D\bar\psi D\psi \exp \left[ - (S_0 + S_1 + S_{int} + S_{Q} +
S_{\bar Q})\right],$$

$$S_0 = \frac{1}{4} \int d^4 x \left( F_{\mu\nu}^a\right)^2,$$
$$S_1 = -i \int d^4 x \bar \psi^f (\hat \partial +m + \hat v + \gamma_5 \hat a + s
+ i\gamma_5 p)^{fg}\psi^g,$$ \be S_{int} = - \int d^4 x \bar \psi^f g \hat A^a
t^a \psi^f .\label{3}\ee

Here $f,g$ are flavor indices, $S_Q$ and $ S_{\bar Q}$  refer to action of
external quark currents, of (possibly high  mass) quark $Q$ and antiquark $\bar
Q$.

We shall follow derivation of \cite{12,13,14}, but for simplicity we shall use
the simplest contour gauge, \cite{14**}  so-called Balitsky gauge \cite{15},
where one can write \be  A_\mu (x) = \int_{C(x)} \alpha_\mu (u) F_{i\mu} (u)
du_i, ~~\alpha_4 =1,~~ \alpha_i = \frac{u_i}{x_i}, \label{4}\ee and the contour
$C(x)$ is going from the point $x= (\vex, x_4)$ to the point $(\mathbf{0},
x_4)$ on the world-line of $Q$ and then along this world-line to $x_4
=-\infty$. Note, that our final result (\ref{10}), (\ref{11})  will be cast in
the gauge invariant form, which is the same  for all contours, connecting
points $x,y$ to the world lines of $Q$ (or $\bar Q$). The independence of the
resulting asymptotic expressions from the form of contours is shown in Appendix
3 of \cite{14}.

 Averaging over fields $A_\mu, (F_{\mu\nu})$, one can write
 \be
Z = \int \int D\psi D\bar\psi  \exp \left[ - (S_1 + S_{eff} )\right],\label{4a}
\ee where $S_{eff}$ was computed in \cite{12}-\cite{14}. Keeping only quadratic
correlators  and colorelectric fields for simplicity\footnote{The fact, that
quadratic (Gaussian) correlators yield dominant contribution is due to small
vacuum correlation length $\lambda\approx 0.1$ fm, which ensures small
expansion parameter $\sigma\lambda^2 \ll 1$, and is strongly supported by
lattice measurements of Casimir scaling (see \cite{10} for discussion . Keeping
only colorelectric correlators is justified when angular momentum of light
quark pair $q\bar q$ is not large. Colormagnetic correlators produce angular
momentum-dependent corrections  $\left(V_{CM} \approx \frac{3l(l+1)}{\sigma
r^3}\right)$ to the linear confinement term $\sigma r$, which is due  to
colorelectric correlators, see \cite{15*} for details.} one obtains (for one
flavor)

\be
 S_{eff} = -\frac{1}{2 } \int d^4 x d^4 y
  \bar \psi   (x)\gamma_4[  \psi (x) \bar \psi (y)]\gamma_4  \psi (y)  J(x,y)
  \label{6}\ee  where $J(x,y)$  is expressed via vacuum correlator of
  colorelectric fields, $[\psi \bar \psi]$ implies color singlet combination.
Keeping only colorelectric fields, one has \be  J(x,y) \equiv \frac{ g^2}{N_c}
  \lan A_4(x) A_4(y)\ran =  \int^x_0  du_i
  \int^y_0 dv_i D(u-v).\label{7}\ee
  Here $D(w)$ is the $np$ correlator, responsible for confinement \cite{10},
  \be \frac{g^2tr}{N_c} \lan F_{i\mu} (u) F_{k\nu} (v) \ran = (\delta_{ik}
  \delta_{\mu\nu} - \delta_{i\nu} \delta_{\mu k} ) D (u-v) + O(D_1)\label{8}\ee
 and we have omitted the (vector) contribution of the correlator $D_1$,
containing gluon exchange  and $np$ corrections to it.

To start one can simplify the matter and neglect possible chiral degrees of
freedom, contained in the $4q$ combination in (\ref{6}), replacing at large
$N_c$ \be  [\psi(x) \bar \psi (y) ]\to  \lan \psi (x) \bar \psi (y)\ran_q
\equiv  S_q (x,y).\label{9}\ee

In this way one obtains \be S_{eff} = \int d^4 x d^4 y \bar \psi (x) \tilde
\mathcal{M} (x,y) \psi(y),\label{10}\ee where $ \mathcal{M}$ is expressed via
$S_q(x,y)$\be \tilde  \mathcal{M} (x,y) =- iS_q (x,y)J(x,y) \label{11}\ee and
$S_q$ is expressed via $\tilde  \mathcal{M}, S_q = \frac{i}{\hat
\partial +m +\tilde\mathcal{M}}$. But one can notice, that $S_q$ contains the
term in the denominator $(m+\mathcal{M})$ and  even for $m=0$ we look for a
scalar $\bar \mathcal{M}$, which provides in $S_q$ terms with even number of
$\gamma_\mu$, and  scalar $\bar \mathcal{M}_s$ can be a self consistent
solution of the  nonlinear equation $\bar \mathcal{M}_s = \frac{1}{\hat
\partial +m+\bar \mathcal{M}_s}J$.  It is clear, that for $m>0$ the situation
is simplified. Below we shall find explicitly the solution of this nonlinear
equation.

At this point one should stress, that $S_q(x,y)$ is the light quark Green's
function in the field of antiquark. To simplify matter, we shall consider
massless quark $(m=0)$ and static antiquark $\bar Q$. It is clear, that without
CSB $S_q$ contains  odd number  number of $\gamma$ matrices, hence  $ tr \tilde
\mathcal{M} (x,y)$ vanishes. To exemplify the appearance of  CSB it is
convenient to consider  the limit of small correlation length $\lambda$ when
$\tilde \mathcal{M}$ factorizes, \cite{12,13} \be \tilde \mathcal{M} (x,y) =
\tilde \delta(x_4-y_4) M(\vex, \vey), ~~ M(\vex, \vey) = J ( \vex, \vey) \beta
\Lambda (\vex, \vey)\label{11a}\ee where $\Lambda(\vex, \vey)$ is expanded in
eigensolutions   $\{\psi_k\}$ of static Dirac equation \be \Lambda(\vex,\vey)=
\sum_{\varepsilon_k} \psi_k (x)~ {\rm sign}~ \varepsilon_k \psi^+_k (y) , ~~ x
\equiv |\vex |.\label{11b}\ee $J(\vex, \vey)$, defined in  (\ref{7}), behaves
at large and parallel $|\vex|\approx |\vey|$ as $$ J(\vex, \vey) \approx {\rm
const}\min (x,y).$$

This implies confinement, if $\Lambda(\vex, \vey)$ ensures  almost equal and
parallel $\vex,\vey$. Hence for selfconsistency $\Lambda(\vex,\vey)$ must tend
to $\beta \delta^{(3)}(\vex-\vey)$ at least for large $|\vex|\sim |\vey|$.
Note, that $\{\psi_k\}$ should be computed with the  interaction $M(\vex,
\vex)$, which consists of the same $\{\psi_k\}$, as it  happens in the mean
field method. And here the spontaneous CSB reveals itself in the fact, that for
scalar mean field $M(\vex,\vex)$ the combination of $\{\psi_k\}$ entering in $
\Lambda(\vex,\vey)$ indeed has this property: in the $4\times 4$ structure
$\Lambda \equiv\left( \begin{array}{ll} \Lambda_{11}&
\Lambda_{12}\\\Lambda_{21} & \Lambda_{22}\end{array} \right)$ one obtains
$\Lambda_{22} =-\Lambda_{11}$ (i.e. $\Lambda\sim \beta)$, if the spectrum
$\{\varepsilon_n\}$ has the symmetry property for $\varepsilon_n \to -
\varepsilon_n$ of the scalar potential and in this case the sum (\ref{11b})
computed in the relativistic WKB   method \cite{16} yields the  smeared
$\delta^{(3)}$ - function at large $x$, ensuring small angle between $\vex$ are
$\vey$ \cite{12}. The appearance of this  scalar structure of $M\sim \beta
\Lambda, \Lambda\sim \beta$ can be called  the spontaneous scalar generation,
which gives CSB.

In the static limit, when one can neglect the energy transfer from the heavy
quark $Q$ to the light quark pair $q\bar q$ in (\ref{10}), the effective mass $
\mathcal{M} (x,y) $ was calculated in  \cite{12,13,14}, using relativistic WKB
approximation\cite{16}. The result is \be \tilde \mathcal{M} (x,y) = \sigma
\left| \vex \right|\cdot \tilde \delta^{(3)} (\vex-\vey) \tilde\delta
(x_4-y_4),\label{12}\ee where $|\vex|, |\vey|$ are distances to $Q$ and  the
smeared $\delta$-function $\tilde \delta^{(3)}(\vez)$ has the range $r_0
=\sigma\left| \frac{\vex+\vey}{2} \right|$, while  the range of $\tilde
\delta(x_4-y_4)$ is of the order of the vacuum correlation length. The explicit
form of $\tilde \mathcal{M}(x,y)$ in (\ref{12}) is given in the Appendix. Hence
for a long string, when $r_0\gg 0.1$ fm, one can neglect the nonlocality in
(\ref{10}) and take into account, that the equivalent contribution can be
obtained, connecting the correlator $J(x,y)$ to the worldline of antiquark
$\bar Q$, which gives the final result  \be S_{eff} = \int d^4 x \bar\psi (x)
\bar \mathcal{M} (x) \psi (x)\label{13}\ee with \be \bar \mathcal{M} (x) =
\sigma (|\vex -\vex_Q| + |\vex - \vex_{\bar Q}|).\label{14}\ee

Two terms of $\bar \mathcal{M} (x)$ are  shown in  Figs. 1 and 2.

There an example of trajectories (world lines) of quarks and antiquarks with
time growing in horizontal direction. Vertical lines starting at $x$ and $y$
belong to  contours $C(x), C(y)$ used in (\ref{4}). Cross-hatched region
between them contributes to the kernel $\tilde M (x,y)$ in (\ref{12}).

 There  are several properties of the interaction (\ref{13}), (\ref{14}) which
 should be mentioned. First of all, the mass term $\bar \mathcal{M} (x)$ is
 scalar and hence creates the $~^3P_0$ pair $\bar q q$.

 Additional terms in $\bar \mathcal{M} (x)$, which are subleading at large
 distances, were studied in \cite{12,13,14} see Appendix below. In particular,  the contribution of
 the term $D_1$ in (\ref{8}), which was neglected above, yields terms proportional to $\beta$  in  $\bar \mathcal{M}
 (x)$, which are of the  relative order $O\left( \frac{\alpha_s}{\sigma r^2}\right),
 r=|\vex-\vex_Q|$, and hence can be neglected at large $r$.

 Secondly,  the kernel  (\ref{14}) does not depend on flavor, which can be checked experimentally, as discussed in concluding section.
 However, the local form (\ref{14}) is an approximation of the nonlocal one, given in
 (\ref{12}) and in Appendix, Eq. (\ref{A1.7}) derived for zero mass $m_q$. The
 effect of nonlocality is numerical reduction of $\bar \mathcal{M} (x)$
 (\ref{14}) by (10$\div 15$)\%, as can be seen from Fig.1 of \cite{13}. For
 larger $m_q$ the range of nonlocality decreases, as shown in Appendix, Eq.
 (\ref{A1.10}), and $\tilde \mathcal{M}(x,y)$ tends to the final form
 (\ref{14}).

  The potential kernel (\ref{14}) has similarity with the kernel, suggested by
  Eichten et al \cite{5}, however, there was assumed its color octet and  vector character,
  i.e. $\bar \mathcal{M} (x)$ was taken to be proportional to $\gamma_4 t_a$, while in our case it is scalar and color singlet.
  This assignment of \cite{5}
  is not  favored  phenomenologically, see \cite{3,4} for discussion. The
  kernel (\ref{14}) coincides formally  with the so-called $sKs$ kernel, which was
  studied nonrelativistically in \cite{6}, and  shown to be phenomenologically successful  and close to  the  $~^3P_0$ model.

  \section{Emission of mesons, accompanying the string breaking process}

  We return here to the $4q$ effecting action (\ref{6}), and will follow the
  procedure of \cite{14*}, where chiral Lagrangian was derived in  conjunction
  with the confining kernel (\ref{12}). Doing bosonization in (\ref{6}) with
  bosonic scalar variable $M_s$ and pseudoscalar $\phi_a$, one obtains the
  effective quark-meson Lagrangian

\be Z = \int D\bar\psi D\psi D M_s D \phi_a \exp \left[ -
S_{QM}\right],\label{15}\ee
$$S_{QM} = - \int d^4 x d^4 y \Bigl[ \bar \psi^f_{ \alpha} (x) \Bigl( i (\hat \partial + \hat v + \gamma_5 \hat a +
s + i\gamma_5 p)^{fg}_{\alpha\beta} \delta^{(4)}(x-y) +
  $$
\be\quad\quad + i M_s (x,y) \hat U_{\alpha \beta}^{fg}(x,y)\Bigr) \psi^g_{
\beta} (y) -  2 N_f \left( J(x,y)\right)^{-1} M_s^2(x,y) \Bigr],\label{16}\ee
\be \hat U_{\alpha \beta}^{fg}(x,y) = \exp\left(i \gamma_5 t_n \phi_n (x,y)
\right)_{\alpha \beta}^{fg}.\label{18}\ee

Integrating out quark fields, one obtains effective chiral Lagrangian given in
\cite{14}  for the field $\phi_a$ with external currents included.

Finally, classical equations of motion define the stationary point conditions,
\be \phi_a^{(0)}(x,y)=0,\label{19}\ee
 \be M_s^{(0)}(x,y)=\frac{-i}{4 N_f } J
(x,y) Tr_{f,d} \left( S_q(x,y) \right),\label{20}\ee \be  S_q(x,y) \equiv
S_{\phi}(x,y)\bigr|_{\phi=0}.\label{21}\ee where $Tr_{f,d}$  is the trace over
 flavor and bispinor indices, $S_\phi (x,y)$ is the total quark
propagator with chiral and extra mesons included, \be S_{\phi}(x,y) =\left\lan
{x} \left|\frac{1}{i \hat \Delta +
    i M_s e^{i \gamma_5 t_a \phi_a} }\right|{y} \right\ran \label{22}\ee
    and  \be  \hat \Delta = \hat \partial + m+\hat v + \gamma_{5 }\hat a +
    s+i\gamma_5  p. \label{23}\ee

    Note, that the equation (\ref{20}) coincides with (\ref{11}), when in $S$
    external  currents are retained. Several properties of the new effective
    action (\ref{16}) are to be noted.

    First of all, chiral degrees of freedom are now contained in the new string
    breaking term, which in the local approximation has the form,
    \be S^{\rm (chiral)}_{\rm str.br} =- i \int d^4 x \hat \psi^f_\alpha (x) \hat M
    (x)Z^{(U)}
    \hat U^{fg}_{\alpha\beta}  (x) \psi^g_\beta (x).\label{24}\ee
    Here $f,g$ are flavor indices, $ \alpha,\beta$ are 4-spinor indices, and
    color indices of $\bar \psi$  and $\psi$ are suppressed, while $\bar M(x) \hat
    U$ are color-blind. The factor
    $Z^{(U)}$  in (\ref{24}) takes into account both np and perturbative renormalization of the $NG$ fields, which are
     nonlocal  and free (without interaction) in (\ref{16}), (\ref{18}),
    while in (\ref{24}) already physical local fields are contained.

\be \hat U =\exp (i \gamma_5 \phi^a t^a) =\exp \left(i \gamma_5 \frac{\varphi_a
\lambda_a}{f_\pi}\right),~~ \varphi_a\lambda_a \equiv \sqrt{2}
\left(\begin{array}{lll}
\frac{\eta}{\sqrt{6}}+\frac{\pi^0}{\sqrt{2}},& \pi^+,& K^+\\
\pi^-,&\frac{\eta}{\sqrt{6}}-\frac{\pi^0}{\sqrt{2}},& K^0\\ K^-,& \bar K_0,
&-\frac{2\eta}{\sqrt{6}}\end{array}\right).\label{25}\ee

In  a similar way the string breaking with emission or absoption of a non - NG
meson is described by the first term in (\ref{16}), see Fig.2

\be S^{(j)}_{str.br} =- i \int d^4 x \bar \psi^f_\alpha (x) \hat
j^{fg}_{\alpha\beta} (x) \psi^g_\beta (x),\label{26}\ee where \be \hat
j^{fg}_{\alpha\beta} (x) = (\hat v + \gamma_5\hat a + s + i \gamma_5
p)^{fg}_{\alpha\beta}.\label{27}\ee

At this point one should stress that the    decay of $Q\bar Q$ meson into
$(Q\bar q) (\bar Q q)$ mesons is possible only  due to string breaking, while
decay into $(Q\bar q ) (\bar Q q) +$ light meson is going via  the string
breaking emission mechanism of Eqs. (\ref{24}), (\ref{26}) and  in addition via
two-step process with string breaking and subsequent emission from light quark
(antiquark) in heavy-light products. When $Q(\bar Q)$ is a light quark, there
is in addition a possibility of emission of $NG$ meson (or light vector meson
etc.) directly from $Q$.\\

\unitlength 0.8mm 
\linethickness{0.4pt}
\ifx\plotpoint\undefined\newsavebox{\plotpoint}\fi 
\begin{picture}(164,63.5)(0,0)
\put(69.5,42.5){\oval(62,29.5)[l]} \put(72.63,42.88){\oval(50.75,11.75)[l]}
\put(136.88,43.38){\oval(54.25,30.25)[l]} \put(123.63,37.13){\oval(.25,.25)[]}
\put(138.75,43.5){\oval(40,12.5)[l]} \put(47.25,57.5){\line(0,-1){14.25}}
\put(52,57.25){\line(0,-1){8}}
\multiput(52,49.25)(.03125,-.0625){8}{\line(0,-1){.0625}}
\multiput(118.75,43.25)(.033333,-.966667){15}{\line(0,-1){.966667}}
\put(124,37.25){\line(0,-1){8.5}}
\multiput(51.75,57.25)(-.0477528,-.0337079){89}{\line(-1,0){.0477528}}
\multiput(47.5,54.25)(.53125,.03125){8}{\line(1,0){.53125}}
\multiput(51.75,54.5)(-.04639175,-.03350515){97}{\line(-1,0){.04639175}}
\put(47.25,51.25){\line(1,0){5}}
\multiput(52.25,51.25)(-.06,-.0333333){75}{\line(-1,0){.06}}
\put(47.75,48.75){\line(1,0){4.5}} \put(52.25,48.75){\line(-5,-2){5}}
\multiput(120.5,39.75)(-.0460526,-.0328947){38}{\line(-1,0){.0460526}}
\put(118.75,38.5){\line(4,-1){5}}
\multiput(123.75,37.25)(-.0913462,-.0336538){52}{\line(-1,0){.0913462}}
\put(119,35.5){\line(4,-1){5}}
\multiput(124,34.25)(-.0865385,-.0336538){52}{\line(-1,0){.0865385}}
\multiput(119.5,32.5)(.1055556,-.0333333){45}{\line(1,0){.1055556}}
\multiput(124.25,31)(-.1111111,-.0333333){45}{\line(-1,0){.1111111}}
\multiput(119.25,29.5)(.1447368,-.0328947){38}{\line(1,0){.1447368}}
\put(119.25,29.25){\line(0,-1){.5}}
\multiput(47.5,56)(.59375,-.03125){8}{\line(1,0){.59375}}
\multiput(52.25,55.75)(-.0579268,-.0335366){82}{\line(-1,0){.0579268}}
\put(47.5,53){\line(1,0){4.75}}
\multiput(52.25,53)(-.0533708,-.0337079){89}{\line(-1,0){.0533708}}
\multiput(47.5,50)(.5625,.03125){8}{\line(1,0){.5625}}
\multiput(52,50.25)(-.054878,-.0335366){82}{\line(-1,0){.054878}}
\multiput(121.25,38.5)(-.075,-.033333){30}{\line(-1,0){.075}}
\multiput(119,37.5)(.1055556,-.0333333){45}{\line(1,0){.1055556}}
\multiput(123.75,36)(-.0791667,-.0333333){60}{\line(-1,0){.0791667}}
\multiput(119,34)(.1111111,-.0333333){45}{\line(1,0){.1111111}}
\multiput(124,32.5)(-.0913462,-.0336538){52}{\line(-1,0){.0913462}}
\multiput(119.25,30.75)(.125,-.0328947){38}{\line(1,0){.125}}
\multiput(124,29.5)(-.184783,-.032609){23}{\line(-1,0){.184783}}
\multiput(119.75,28.75)(.03125,-.0625){8}{\line(0,-1){.0625}}
\put(52.75,62.5){\makebox(0,0)[cc]{$ \bar Q$}}
\put(58.5,46.5){\makebox(0,0)[cc]{$q$}} \put(54.75,23){\makebox(0,0)[cc]{$ Q
$}} \put(58,39.5){\makebox(0,0)[cc]{$\bar q$}}
\put(44,42.25){\makebox(0,0)[cc]{$x$}} \put(55.25,51){\makebox(0,0)[cc]{$y$}}
\put(123.25,63.5){\makebox(0,0)[cc]{$ \bar Q $}}
\put(130,47.25){\makebox(0,0)[cc]{$q$}} \put(115.25,42){\makebox(0,0)[cc]{$x$}}
\put(126.5,40.25){\makebox(0,0)[cc]{$y$}}
\put(134.5,40.25){\makebox(0,0)[cc]{$\bar q$}}
\put(130.75,24){\makebox(0,0)[cc]{$ Q $}} \put(47,43){\circle*{1.8}}
\put(52,49){\circle*{1.5}} \put(118.5,43){\circle*{2.24}}
\put(124,37.75){\circle*{1.5}} \put(45.75,14){\makebox(0,0)[cc]{F i g. 1}}
\put(117.5,14.25){\makebox(0,0)[cc]{F i g. 2}}

\end{picture}

\section{Matrix elements of string decay and hadron emission}

In this section we shall study physical matrix elements based on effective
actions (\ref{13}), (\ref{24}) and (\ref{26}).

The basic quantity, which can be calculated from the effective Lagrangians,
derived above, is the two-step amplitude, which incorporates string breaking
transition to intermediate two-body states and back to $Q\bar Q$ state.

In this way the transition amplitude from the state $(Q\bar Q)_n$  to $(Q\bar
Q)_m $ via string  breaking to the intermediate states $(Q\bar q)_{n_2}(\bar Q
q)_{n_3}$ with energy $E_{n_2n_3}$ can be written as (see \cite{18,19,20} for
details, a similar amplitude was introduced in \cite{5})

\be w_{nm} (E) = \frac{1}{N_c} \int \frac{d^3\vep}{(2\pi)^3} \sum_{n_2, n_3}
\frac{\tilde J_{nn_2n_3} \tilde J^+_{mn_2n_3}}{E- E_{n_2n_3}
(\vep)}.\label{29}\ee Here $\tilde J_{nn_2n_3} (\vep)$ is the overlap matrix
element with $\bar M(\vex) \to M(\veq)$

\be \tilde J_{nn_2n_3} (\vep) = \int \bar y_{123} \frac{d^3\veq}{(2\pi)^3}
\Psi^+_n (\vep+\veq)\bar M(\veq)  \psi_{n_2} (\veq) \psi_{n_3}
(\veq)\label{30}\ee and $\bar y_{123}$ is the trace of normalized spin-tensors
corresponding to spin-angular parts of meson states, while $\Psi_n, \psi_n$ are
the radial parts, tables of $\bar y_{123}$ are given in \cite{18,20},e.g. for
the $1^{--}$ state $n$ one has $\bar y_{123} =\frac{i}{\omega} (q_i
-\frac{p_i\omega}{2(\omega+\Omega)})$ where $\omega,\Omega$ are average kinetic
energies of light and heavy quark respectively, tables of  $\omega,\Omega$ are
given in Appendix 1 of \cite{18}.

For the string breaking operator (\ref{14}) the overlap integral (\ref{30}) has
the form \be \tilde J_{nn_2n_3} (\vep) = \sigma \int \bar y_{123}
\frac{d^3\veq}{(2\pi)^3} \Psi^+_n (\vep +\veq) \left(\left|\frac{d\psi_{n_2}
(\veq)}{d\veq}\right| \psi_{n_3} (\veq)+ \left|\frac{d\psi_{n_3} (\veq)}{d\veq}
\right|\psi_{n_2} (\veq)\right)\label{31}\ee

Since heavy-light meson wave functions are well reproduced by a single
oscillator function \cite{18,19,20}, $$ \psi_{n_2} (\veq) = \left(
\frac{2\sqrt{\pi}}{\beta_2} \right)^{3/2} e^{-\frac{\veq^2}{2\beta^2_2}},$$ the
resulting effective  vertex $\bar M(\veq)$ can be written as \be \bar M_{\rm
eff}(\veq) =  2\sigma \frac{\lan q_{\rm eff}\ran}{\beta_2^2}\approx
\frac{2\sigma}{\beta_2} \equiv M_\omega\label{32}\ee

This constant parameter $M_\omega$ was systematically used in \cite{18,19,20}
to calculate decays of bottomonium. In this case $\beta_2(B) =0.48$ GeV and
$M_\omega\approx  0.8 GeV;$ this value was exactly used in \cite{18,19,20}, and
one can see, that it agrees with the vertex mass operator $\bar M(x).$

Using $w_{nm}$, one can calculate production and  decay of all states involved.

\section{Comparison to $~^3P_0$ and $jKj$ models}

Our derivation of the effective Lagrangian (\ref{14}) implies,  that the basic
interaction behind the string breaking is the scalar confining term acting
between the light and heavy (anti)quark, (with subdominant terms of vector
color singlet OGE type etc.) This is in exact correspondence with the so-called
$sKs$ model, studied in detail  in \cite{6} and compared to the $~3P_0$ model,
where it was shown that both are successful and give similar results. In this
way our derivation of $\bar M(|\vex|)$ in (\ref{14}) gives additional
theoretical status for both  $sKs$  and $~^3P_0$ models as candidates for the
string breaking kernel. The difference between our approach and $sKs$ model  in
\cite{6} lies in the treatment of quark motion in the decay matrix elements,
which was taken nonrelativistic in \cite{6}, with constituent quark mass $m_q$,
while in our approach we are using relativistic formalism for light quarks with
current (zero) mass. In particular, our formalism for calculation of the factor
$\bar y_{123}$ via vertex $Z_i$ factors in Appendices 1 and 2 of \cite{20} and
the use of relativistic string Hamiltonian for $(Q\bar Q)_n$ and $(Q\bar q)_n$
states, takes into account relativistic kinematics. With all that difference,
results of our analysis are qualitatively and even quantitatively   similar to
those of $sKs$ and  $~^3P_0$ models.

In particular, the width and shift of the state $(Q\bar Q)_n$ is given by \be
(\Delta E_n, \Gamma_n) = (Re w_{nn}, -Im w_{nn})\label{34}\ee and for the decay
$\Upsilon(4S) \to B\bar B$ one obtains \be  \Gamma_{4S}(B\bar B)= \left(
\frac{M_\omega}{2\omega}\right)^2 0.0033~ |J_{B\bar B} (p)|^2 ~({\rm
GeV)}.\label{37a}\ee One can now estimate $M_\omega$ from the width of the
decay $\Upsilon(4S) \to B\bar B$, as it was done in \cite{19}. The wave
functions of all participants have been found with good accuracy from the
relativistic string Hamiltonian \cite{22*} and parametrised by 15 oscillator
functions. Comparing experimental value $\Gamma_{\exp} (\Upsilon(4S) \to B\bar
B)= (20.5\pm 2.5) $ MeV \cite{22} with (\ref{31}) one obtains $M_\omega\cong
0.8$  GeV. Another check of our kernel (\ref{14}) and its average value
$M_\omega$ is the decay of $\psi(3770)$ into $D\bar D$ with the width
$\Gamma_{\exp} \cong 25.4$ MeV \cite{22}. Following \cite{22*}, one takes this
state as $1^3D_1$ and approximates it with 5 oscillator functions. Calculating
(\ref{30}) with $\bar M(q) \to M_\omega=0.8$ GeV, one obtains $\Gamma_{th} =22$
MeV, which agrees with $\Gamma_{\exp} $ within experimental $(\sim 10\%)$
accuracy\footnote{The author is grateful to V.D.Orlovsky, who provided detailed
calculations of this decay (to be published) .}.

 One can now compare this
value with (\ref{14}) and take into account, that the r.m.s. radii for
$\Upsilon(4S)$ and $B$   are  0.9 fm  and 0.5 fm \cite{22*} respectively.
Therefore, one obtains $\sigma  2 r_B\approx 0.9$ GeV, and $\sigma r_\Upsilon
\approx 0.81$ GeV, in good agreement with (\ref{32}).

 In a similar way  the decay of $\Upsilon(5S)$ into
 six channels of $B_i \bar B_k$, where $B_i(B_k)= B, B^*, B_s, B_s^*$ was considered in \cite{23}.
 The total width computed in \cite{23} is $\Gamma_{\rm tot}= 116$ MeV $ M^2_\omega$,
 where $M_\omega$ is  in GeV. Comparing to $\Gamma_{\rm tot} (\exp) = (110 \pm
 13)$ MeV \cite{22}, one has $M_\omega=(0.91\div 1.03)$ GeV in reasonable
 agreement with (\ref{32}).
 Also the experimental ratio of  decay into beauty-strange mesons is well reproduced
 in \cite{23}, which supports the flavor-blind   kernel $\bar M$ in (\ref{11}).

 We now turn to the string-breaking emission process. It is best studied in the
 transitions of the type $(Q\bar Q)_n \to (Q\bar Q)_{n'} \pi\pi$
 \cite{18,19,23} and in $(Q\bar Q)_n \to (Q\bar Q)_n \eta $ in \cite{24}.

 In \cite{18,19} the subthreshold string breaking was considered for the decay
 of $\Upsilon(4S)$, while in \cite{23} the decay of $\Upsilon (5S)$ was analyzed, In all cases of this
 double-pion or single eta
 $(n, n')$ transitions the effective $Z^{(U)}$ factor appears to be small,
 $Z^{(\pi)} \approx Z^{(\eta)} \approx O\left(\frac{f_\pi}{M_\omega}\right)$.

In conclusion, we have derived the relativistic string breaking
kernel, which is the scalar color-singlet confining interaction,
which is flavor blind and nonlocal for zero mass $q\bar q$ pair,
tending to the confining $\sigma r$ potential for long breaking
string. This form is close to $~^3 P_0$ and $sKs$ models in the
nonrelativistic formalism. The author is grateful to D.V.Antonov
for many useful remarks and suggestions. Financial support of RFBR
grant no.09-02-00 620a is gratefully acknowledged.

\vspace{2cm}

{\bf Appendix  }\\

{\bf Properties of the mass kernel, Eq.(17)}\\

 \setcounter{equation}{0} \def\theequation{A.\arabic{equation}}

To simplify calculations one can choose the  correlator $D(z)$ in the Gaussian
form, $D(z)= D(0)\exp \left( - \frac{z^2}{4\lambda^2}\right)$, where $\lambda$
is the vacuum correlation length, $\lambda\approx (0.1 \div 0.2)$ fm,  then
$\sigma=\frac12 \int  D(z) d^2 z= 2\pi D(0)\lambda^2$. In the final expressions
for the asymptotics  of $\mathcal{M}(\vex, \vey)$ only $\sigma $ enters,
therefore the value of $\lambda$ and the form of $D(z)$ influences only the
small $x,y$ region (the  same is true for the  dependence on the  contours of
the  contour gauge, see Appendix 3 of \cite{14}).   Now the  mass operator of
Eq. (\ref{11a}) has the form \be M(p_4 = 0, \vex,\vey) \equiv M (\vex, \vey)=
\sigma (\vex\vey) f (\vex, \vey) \beta \Lambda (\vex, \vey),\label{A1.1}\ee
where \be f(\vex,\vey) =\frac{1}{2\sqrt{\pi} \lambda} \int^1_0 ds \int^1_0 dt
\exp [-(\hat x s - \hat y t)^2], \hat x, \hat y =\frac{1}{2\lambda}
(x,y)\label{A1.2}\ee with asymptotics $f(\vex, \vex) = \frac{1}{|\vex|},
|\vex|\to \infty$, and \be \Lambda (\vex, \vey) = \sum_n \psi_n (\vex) sign
\varepsilon_n \psi^+_n (\vey),\label{A1.3}\ee while $\psi_n(\vez)$ satisfy
equations \be \left( \frac{\veal}{i} \frac{\partial}{\partial\vez} + \beta
m\right) + \beta \int \mathcal{M} (\vez, \vew) \psi_n (\vew) d \vew =
\varepsilon_n \psi_n(\vez). \label{A1.4}\ee

The mean-field-type equations (\ref{A1.1}), (\ref{A1.4}) have been solved in
\cite{12}--\cite{14}, using relativistic WKB method \cite{16}, and below are
listed expressions for the asymptotics of $M(\vex, \vey)$, extracted from
\cite{14}, $M(\vex, \vey)$ is  a $4\times 4$ matrix, which can be represented
in the $2\times  2$ form, with entries expressed via Pauli matrices $\sigma_i$,
\be \mathcal{M}_{ik} (\vex, \vey) = \left(\begin{array}{ll} \mathcal{M}_{11}&
\mathcal{M}_{12}\\ \mathcal{M}_{21}& \mathcal{M}_{22}\end{array}\right), ~~
\mathcal{M}_{ik}=a_{ik} \hat 1 + \veb_{ik} \vesig.\label{A1.5}\ee

In the region where $\cos \theta$ between $\vex, \vey$ is close to one, and
$|\vex|\sim |\vey|,$ one obtains \be a_{11} = \sigma |\vex|\Delta_1  (x,y,
\theta) , \veb_{11} = - \frac{\veL\sigma^2}{4\pi} \Delta'_1
(x,y,\theta)\label{A1.6}\ee
$$ a_{12}= \frac{\sigma^2}{2\pi}  \Delta_{12}, ~~ \veb_{12} =
\frac{\sigma^2}{2\pi} (\ven \Delta'_{12} + (\ven \times \veL)
\Delta_{12}^{\prime\prime}.$$

The symmetry of $ \mathcal{M}_{ik} $ is :  $\mathcal{M}_{22}= \mathcal{M}_{11},
\mathcal{M}_{21}= -\mathcal{M}_{12}$.

Here ($|\vex | \equiv  x, ~|\vey|\equiv y)$ \be \Delta_1 (x,y, \theta) =
\frac{\sigma^2x^2 K_1(\sigma {x} \sqrt{(x-y)^2+ \theta^2 x^2})}{2\pi^2
\sqrt{(x-y)^2 + \theta^2x^2}},\label{A1.7}\ee

\be \Delta_1' (x,y, \theta) = K_0 \left(\sigma xy
\sqrt{\theta^2+\frac{(x-y)^2}{xy}}\right).\label{A1.8}\ee

Note, that $\Delta_1$ plays the role of smeared normalized  function $\tilde
\delta^{(3)} (\vex-\vey):$ \be \int \Delta_1 (x,y, \theta)
d^3y=1,\label{A1.19}\ee where one is using relation $\int\delta(1-\cos \theta)
d\cos \theta =\frac12$.

All functions $\Delta_{12}, \Delta_{12}',\Delta_{12}^{\prime\prime}$ are
antisymmetric in $x,y$ and therefore are zero in the local limit $x=y$,
therefore they are not given here, see \cite{14}.

Expressions (\ref{A1.7}), (\ref{A1.8}) are calculated for the case $m_q=0$,
however the derivation in \cite{12, 14} is valid also in the case $m\neq 0$,
where  one can approximately at large $x$ replace $\sigma x$ in (\ref{A1.7}) by
$\sigma x + m_q$, so that  $\Delta_1$ becomes \be \Delta_1 (x,y,\theta) \to
\frac{(\sigma x + m_q)^2 K_1 ((\sigma x + m_q) \sqrt{(x-y)^2+ \theta^2
x^2})}{2\pi^2\sqrt{(x-y)^2 + \theta^2 x^2}}.\label{A1.10}\ee

One can see in (\ref{A1.10}), that the range of nonlocality,  $|x-y|_{eff}$, in
the limit of large $m_q$  tends to zero, and one can   replace $\Delta_1$ by
$\delta^{(3)}(\vex-\vey)$ This leads to a moderate increase  of effective
confinement with growing $m_q$.  The  same sort of mass dependence of confining
interaction between quark and antiquark was observed recently on the lattice
\cite{25}.

\end{document}